\documentclass[conference]{IEEEtran}
\IEEEoverridecommandlockouts
\usepackage{cite}
\usepackage{amsmath,amssymb,amsfonts}
\usepackage{algorithmic}
\usepackage{graphicx}
\usepackage{textcomp}
\usepackage{xcolor}
\usepackage{hyperref}
\hypersetup{
	colorlinks=true,
	linkcolor=blue,
    filecolor=magenta, 
    urlcolor=cyan
}
\def\BibTeX{{\rm B\kern-.05em{\sc i\kern-.025em b}\kern-.08em
    T\kern-.1667em\lower.7ex\hbox{E}\kern-.125emX}}
\begin{document}

\title{A Survey on Serverless Computing}

\author{\IEEEauthorblockN{Jacob John}
\IEEEauthorblockA{\textit{School of Computer Science and Engineering (SCOPE)} \\
\textit{Vellore Institute of Technology}\\
Vellore, India \\
jacob.john2016@vitalum.ac.in}
    \and
\IEEEauthorblockN{Shashank Gupta}
\IEEEauthorblockA{\textit{School of Computer Science and Engineering (SCOPE)} \\
\textit{Vellore Institute of Technology}\\
Vellore, India}
}

\maketitle
\begin{abstract}
The Internet is responsible for accelerating growth in several fields such as digital media, healthcare, the military. Furthermore, the Internet was founded on the principle of allowing clients to communicating with servers. However, serverless computing is one such field that tries to break free from this paradigm. Event-driven compute services allow users to build more agile applications using capacity provisioning and a pay-for-value billing model. This paper provides a formal account of the research contributions in the field of Serverless computing. 
\end{abstract}

\section*{Drawbacks of FaaS}
Baldini et al \cite{b1} presents a formulation of the serverless trilemma formed by three constraints:

\begin{itemize}
\item{Solutions should not violate the black box constraint, i.e., should not follow a fusion structure}
\item{Function composition should obey a substitution principle with respect to synchronous invocation}
\item{Solutions should not violate the double billed constraint, e.g. the same function that forms a sequential shouldn’t be invoked more than once}
\end{itemize} 

Furthermore, the paper also discusses the impact on this trilemma on programming model choices. The paper also presents a model capability of building new serverless functions based on the composition of existing ones. According to Westerlund et al \cite{b2}, function composition problems remained unresolved. The event-driven or reactive core of serverless programming is not sufficiently expressive to implement serverless functions as a composition of functions. Other drawbacks of FaaS include runtime constraints and state constraints \cite{b3}. A developer cannot assume that the state of a function is persistent across multiple invocations. Thus, persistent of the state has to externalized outside the serverless function instance. 

\section*{FaaS for networks}

Cicconetti et al \cite{b4} presents a distributed delegated architectural framework in a Software Defined Networking (SDN)-enabled edge computing domain. This framework allows lamda function requests from mobile client notes to be routed to specific edge devices. It implements a computation and forwarding method, where computation is the runtime environment of the function while forwarding is transferring the function request outside of the edge domain. Since networking devices participate in computation of the lamda functions, it follows a serverless paradigm.

In addition to this, a performance evaluation of the framework is conducted using an edge computing emulator. Experiments were performed using synthetic scenarios adopted from those proposed by Sonmez et al \cite{b5}. Figure 1 illustrates some of these synthetic topologies along with those used by Cicconetti et al. Three scenarios were presented, each with different traffic conditions, lamda request patterns and network topologies. In conclusion, the architecture is capable of handling network conditions and fasting changing loads. This is best achieved by the RR algorithm. A future enhancement for this study could be to implement this architecture in a 5G enabled network. Due to the fast and dynamic nature of mobile edge devices, this architecture would be suitable with the RR algorithm. 

\begin{figure}[htbp]
\centerline{\includegraphics[width=0.3\textwidth]{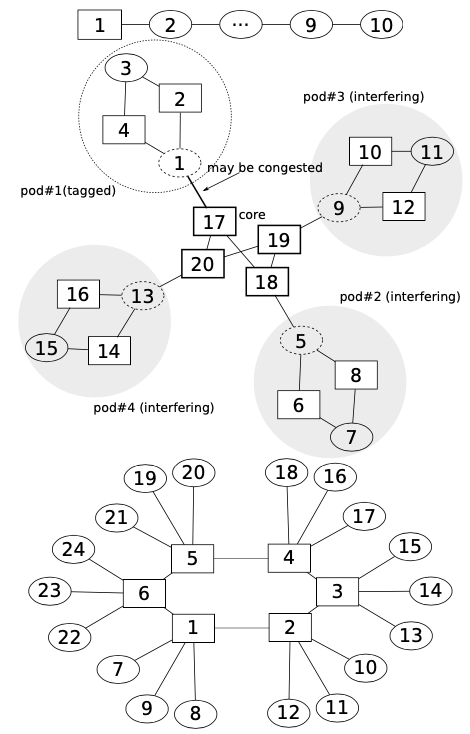}}
\caption{Top – line topology; Middle – pod topology; Bottom – ring tree topology. Rectangles – e-computers; Solid boundary ellipses – e-routers; Dashed boundary ellipses – pure switches.\cite{b4}}
\label{fig1}
\end{figure}

\section*{Snafu}

Spillner presents Snafu \cite{b6} or Snake Functions as an open source and modular system that grants requests for named functions. Clients contain majority of the application logic and are used to externally trigger systems offering stateless micro-services. These systems contain ‘fine-granular’ functions that are invoked via network protocols. TKról et al, terms this concept as Named Function as a Service (NFaaS). The Snafu architecture also allows local invocation of functions via a command-line interface.

Snafu is said to be suitable for quick prototyping of applications and/or ad-hoc experiments. Snafu offers flexibility, modularity and extensibility to balance isolation and authentication. The Snafu architecture realizes this modularity through ``an extensible system whose functionality is partly contained in subsystems with pluggable components". In addition to this, a zero-configuration default setup makes the system comparatively easier to use than AWS Lamda compatible and incompatible runtimes that require integration with other runtime components and frameworks. In addition to this, Snafu also provides the added benefit of, ``per-instance, per-tenant and per-function configuration options (in order of priority)".

A serverless real-time data processing app tutorial given on Amazon’s website \cite{b8} requires the following components to build:

\begin{itemize}
\item{AWS Lambda – to process serverless real-time data streams}
\item{Amazon Kinesis – to archive raw data to S3}
\item{Amazon S3 – to store raw data}
\item{Amazon DynamoDB – to persist records in a NoSQL database}
\item{Amazon Cognito}
\item{Amazon Athena – to run ad-hoc queries against raw data}
\item{AWS IAM}
\end{itemize}

Note that all components belong to Amazon, including S3 that offers storage for the application. Thus, utilizing AWS Lamda might result in a vendor lock-in as all the data is archived and processed on Amazon’s services. 

The worst in-process single threaded performance of Snafu is still competitive to AWS Lamda with a 377.75\% and 188.09\% relative speed for the computation of recursive Fibonacci of 15 implemented using Python. Spillner proposes a parallelization optimization would yield better results and could be implemented in future versions. Snafu also said to achieve better utility than AWS Lamda for more frequent use which makes it more economical while still attaining competitive performance results. However, with large instance types, Snafu gets ``prohibitively expensive" as at some point, the performance halts but the price still climbs.

\section*{McGarth.NET}

McGarth et al [9] develop an open source, performance oriented serverless platform deployed in Microsoft Azure and implemented in .NET. For data persistence and for its messaging layer, this platform relies on Azure Storage Services. The function execution environments are Windows containers. The platform comprises of two primary components – web services, which contains web apps, load balancers, etc. and the worker services. For simplicity, this paper refers to this platform as McGarth.NET. 

McGarth.NET was tested against open source performance tools developed by McGarth et al. The tool utilizes a “Serverless Framework to deploy Node.js functions to different services” [10]. Two tests were conducted – Concurrent Test, measures the ability to performantly scale and execute a function and Backoff Test, to study cold start times and expiration behaviors of function instances. The results for both tests are encouraging, even when compared to commercial services such as AWS lamda, Google Cloud Functions, Apache OpenWisk and Azure Functions. Furthermore, it should be noted that though Azure Functions and McGarth.NET have different function execution environments despite both being Windows implementations. 

Some of the limitations of McGarth.NET include Function immutability – as REST API only supports creation, deletion and read operation of function resources. McGarth et al propose that this could be overcome by storing multiple versions of a function and target each version during execution. Another limitation would be the use of warm queues as a FIFO implementation. This causes issues with container expiration. Furthermore, Azure doesn’t support LIFO queues, but a Redis \cite{b11} cache could be used ``to store warm stacks and use the list data type that supports LIFO queue operations". It should also be considered that Redis and Azure are very different systems in terms of purpose and implementation. 

Asynchronous Execution and Worker Utilization are also other limitations in McGarth.NET. However, this can be overcome using services such as Amazon SDK (Boto3) and Google Functions allow asynchronous, non-blocking invocations via various methods \cite{b12}. Furthermore, since McGarth.NET utilizes windows containers as a substitute for linux containers, certain useful operations such as container resource updating and container pausing \cite{b13} aren’t supported.

\section*{Pipsqueak and PipBench}

Oakes et al introduce an approach called Piqsqueak for larger FaaS functions. By introducing a package caching mechanism, Pipsqueak reduces initialization overhead and speeds up function deployment. This is because modern applications are dependent on userspace dependencies \cite{b16} and large libraries \cite{b15}, such as numpy and scikit present in the PyPI repository. Pipsqueak achieves this by utilizing containers with predeployed Python libraries. Thus, ``FaaS function package sizes can be smaller which enables deployment to be more agile" \cite{b17}. Furthermore, Oakes also proposed PipBench, a tool for evaluation package support.

\section*{ServerlessOS}

Al-Ali et al \cite{b18} propose a novel method of abstraction for serverless computing. This abstraction allows for the support of general rather than solely event driven processes. ServerlessOS comprises of a multi-resource disaggregation model, a fine-grained cloud orchestration layer and an isolation capability. The coordinated isolation (among workloads) extends the Linus cgroup functionality and enables data privacy and resource management across disaggregation.

Al-Ali also claim to provide the ``same abstraction developers are familiar with." Thus, allowing legacy code to be redeployed without having the need to refactor it. Furthermore, since ServerlessOS is a process-based model that constantly evolves to provide many customized features, it avoids vendor lock-in for consumers. In addition to this, due to its resource disaggregation properties, this serverless model abstracts resources as a `pool of CPU, I/O and memory" which helps fulfill growing demands.

\section*{Recycle.io}

Al-Masri et al \cite{b19} utilize serverless and edge computing to create a smart waste management system, Recycle.io \cite{b20}. This novel device reduces the cost associated with separation of waste by providing waste management agencies with smart recycling and organic bins. Detection happens at the edge of the IoT network using Azure functions that utilize data captured from a camera and ultrasonic sensor module. Furthermore, the serverless functionality is the use of Microsoft’s Custom Vision to label images as sources of waste. Violations at smart bins are captured and stored into an SQL database on the cloud and can be monitored using a dashboard.

\section*{IBM Bluemix OpenWisk}

OpenWisk (proposed by Baldini et al \cite{b21}) is an open source serverless mobile application development platform used for constructing cloud native actions. By supporting multiple languages and APIs, OpenWisk allows customization and hence simplifies development of mobile application architectures. This coupled with the availability as open source \cite{b22} prevents vendor lock ins. Furthermore, the support of OpenWisk packages allows developers to share various implementations.

An event-based programming model similar to ECA \cite{b23} is used develop trigger/action/rules for the model. A composition of multiple actions forms an action sequence. OpenWisk also unloads resources on-demand basis each invocation and trigger and uses Docker containers to provide auto-scaling. Thus, allowing developers to focus primarily of application development rather than provisioning and maintaining cloud-based APIs. In addition to this, Baldini et al present a Weather and Watson configuration that uses the Weather.com API to retrieve the weather forecast for 10 days. Followed by a translate action that translates the forecast from English to Spanish.

\section*{The freeze/thaw cycle}

According AWS’ official documentation [24], after configuring an environment and launching it, a container based on the execution context specified in the configuration is created. AWS Lambda provisions and manages the configured resources such as amount of memory and maximum execution time. AWS lambda also maintains an execution context in the cache, hence reducing the initial overhead.  These executing contexts are reused for subsequent invocations to prevent the added latency and bootstrapping every time successive lambda functions are invoked. Furthermore, in Node.js, every function will be marked complete in one of the following ways \cite{b13}:

\begin{itemize}
\item{\textbf{Timeout}: The maximum execution time is specified by the user prior to container creation. This initialization code is run before any calls are made or the event handler is fired for the first time. Once this user-defined duration is availed, the function or the code in the container will ``summarily halted” regardless of the state the function is in. Hence, it should be noted that the user should perform dry runs on the code to test what time threshold would yield an undesired exception for the application. For example, recursive loops with poor exit parameters set would relentlessly lead to abnormal behavior.}
\item{\textbf{Controlled termination}: If any callback invokes \textit{context.done()} and then finishes its own execution, the remaining callbacks would also terminate regardless of their execution state. It should be noted that this callback doesn’t necessarily need to be the ``original handler entry point”. The user should also note that explicit termination should not impact the performance of other callbacks.}
\item{\textbf{Default termination}: If all processing callbacks complete, then the container will terminate. A ``Process exited before completing request” message is written to the logs. This occurs when no \textit{context.done()} is called upon completion of the terminal or final callback. The user shouldn’t perceive the log message as an error but rather an implicit termination of the running code.}
\end{itemize}

A \textit{freeze/thaw} cycle is employed in order to maintain an execution context or ``keep-alive” a container. Freeze refers to the process of preserving the directory content when a function terminates. This resides in the \textit{/tmp} directory of 512 MB which is inherent to every execution context. Hence equipping the client with a transient cache or ephemeral disk space that is utilizes in any preceding lambda invocation. Despite the process being automated, AWS suggests, ``adding logic in your code” to examine whether a connection persists prior to requesting a fresh connection.

An intrinsic property in almost all FaaS platforms is autoscaling and deprovisioning. This is done in order to conserve server capacity, energy and reduce costs for clients. Granular billing is utilized and hence the user is billed when the serverless application is being execute \cite{b25}. However, this is done by allowing the hosting infrastructure to go COLD. This means rather than running the containers constantly, to conserve server resources, which can be then harnessed by others, service providers typically deprovision server capacity when demand for it falls. 

Lloyd et al discuss the implications of this deprovisioning and its impact on microservice performance via a comprehensive investigation \cite{b26}. In this paper, the authors forced cold runs by creating new HTTP-triggered function apps containing single C\# functions for each run. The behavior of each function was measured on the Microsoft Azure platform using Visual Studio Team System (VSTS). Performance load tests and stress tests were performed on the functions using VSTS \cite{b27}. 

This paper also terms four states of serverless infrastructures. Each of these serverless microservice container states are run in Docker \cite{b28}\cite{b29}. Docker containers also support specification of CPU-period. A default CFS or Completely Fair Scheduler is the default Linux kernel CPU scheduler for normal Linux processes. Runtime flags can be used to configure the CPU scheduler period in 100 micro-seconds intervals. The \textit{ --cpu-period=<value>} can be used to change this value.

\begin{itemize}
\item{\textbf{Provider cold}: First function invocation made to the cloud provider.}
\item{\textbf{VM cold}: First function invocation made to the virtual machine.}
\item{\textbf{Container cold}: First function invocation made to container hosting microservice code. Containers run directly the operating system kernel and each workload has access to a restricted subset of resources.}
\item{\textbf{Warm}: Repeating function invocation of code in transient cache or in preexisting container. This is analogous to the AWS’ freeze/thaw cycle that maintains an execution context even after the callback terminates its execution.}
\end{itemize}

To emulate TCP networking overhead incurred by the AWS API Gateway and Lambda function invocation, Docker was used in the experimentation process. However, during remote tests, minimal load was observed on the client. Cold and warm performances on Lambda and AWS were compared with Docker-Machine performance. Warm performance on Lambda was reasonable compared to equivalent implementation using remote Docker containers to host code. 

Lloyd et al also discussed the impact of memory reservation on container placement and its implications on performance. When increasing from 128 MB to 256MB of RAM, AWS allocates twice as much CPU power to lambda functions \cite{b32}. Furthermore, it should be noted that increasing RAM to 3 GB or 3,072 MB doesn’t necessarily imply that 24 CPUs are dedicated to the function. This would be both uneconomical and introduces an added layer of complexity to run a client’s code in a parallel and effective manner. 

Based on the results in Lloyd et al’s experiment, COLD service execution time experienced a 4 times performance boost on average. The memory reservation, in this case, was increased from 128MB to 1,536MB, which is a twelve-fold increase. Hence, further establishing the fact that CPU performance doesn’t scale strictly proportionally with memory reservation \cite{b33}. On the other hand, only a four-fold performance improvement was observed in WARM service execution time. Furthermore, performance improvements slowly diminish in WARM infrastructures beyond 512MB memory reservation. 

The containers took 10 minutes to go “cold” or be deprecated, followed by VMs. 40 minutes later, microservices, VMs and containers are no longer in use on the hosting infrastructure. This COLD infrastructure suffered from a 15x performance degradation after 40 minutes of inactivity. However, in order to counteract the initialization overhead incurred on startup during COLD function invocation requests, some platforms provision additional infrastructures.

Pérez et al \cite{b34} discuss the impact of increasing container size and the time it takes for the container to cached in the ephemeral disk space. For example, 30 invocations are required to cache a c7 function which is said to have a medium container size. This has an image size of 70 MB and uses Docker image centos:7 \cite{b35}. This is compared with the regards to largest container being utilized, ub14. This has an image size of 153 MB, more than twice as much as the mid-sized container. Furthermore, it uses the uses the Docker image \textit{grycap/jenkins:ubuntu14.04-python}. This container requires approximately 80 invocations, more than twice as much, until cached. 

Since provision is slow due to the use of elastic load balancing schemes in services such as AWS’ elastic load balance \cite{b37}.  Such services utilize the Load Balancer as a Service (LBaaS) model but are designed to respond to current or future service demands \cite{b38}. This yields poor execution and deployment times for VMs that require additional initialization beyond OS bootstrapping. A solution for this performance degradation due to COLD infrastructures was suggested by Llyod et al \cite{b17}. A recommended mechanism to keep the container warm is to set clients to trigger serverless functions with a set delay. This allows a desired amount of concurrent executions to build up thus mitigating cold start latency as the infrastructure is preserved \cite{b39}. 

A use of a 100 concurrent client requests are employed to keep-alive workloads and preserve serverless performance after idle periods. This prevents automated deprovisioning of computing power in the form of resources. Recycling infrastructure on the serverless platform happens in the form of transient cache or subsequent calls to function containers \cite{b26}. Thus, preventing the overhead incurred due to reinitialization of cold containers or cold VMs. Furthermore, it prevents the added latency induced to service response times due to future invocations that force initialization of new server infrastructure. The results were highly favorable and yielded approximately 18x cost savings. Furthermore, hosting infrastructure was retained during the freeze/thaw cycle by leveraging keep-alive workloads. Thus, also increasing speedups with reasonable average runtimes for each workload.

\section*{Exploring Serverless Computing for Neural Network Training}

Lang Feng \cite{b40} research the utilization of serverless runtimes while utilizing information parallelism for expansive models, demonstrate the difficulties and restrictions because of the firmly coupled nature of such models and propose changes to the hidden runtime usage that would moderate them. For hyper-parameter advancement of littler profound learning models, they demonstrate that serverless runtimes can give a noteworthy advantage.

\section*{Visualizing Serverless Cloud Application Logs for Program Understanding}

Chang and Fink \cite{b41} present a tool that imagines cloud execution logs for an alternate objective – to encourage program understanding and create documentation for an application utilizing runtime information. Our device presents another timetable perception, another strategy, and UI to abridge various JSON articles and present the outcome, and cooperation procedures that encourage exploring among capacities. Together, these highlights clarify a serverless cloud application's piece, execution, dataflow and information construction. We report some underlying client input from a few master engineers that were engaged with the apparatus' structure and improvement process.

They built up a tool called Witt that imagines a serverless application's execution logs for an alternate reason – to enable designers to see how the application is developed utilizing runtime information. An undertaking application is regularly made in a community domain including numerous designers \cite{b9}.

\section*{Serverless Programming (Function as a Service)}

Castro and Ishakian \cite{b42} discusses Serverless Computing (otherwise called Function as a Service) as a emerging as another and convincing worldview for the sending of cloud applications, to a great extent because of the ongoing movement of big business application structures to compartments and microservices. They clarify from the point of view of a cloud supplier, serverless figuring gives an extra chance to control the whole improvement stack, decrease operational expenses by productive streamlining and the board of cloud assets, and empowering a serverless biological system that supports the organization of extra cloud administrations

\section*{Serverless Computing: Economic and Architectural Impact}

Gojko Adzic and Robert Chatley \cite{b43} presents two case modern investigations of early adopters, demonstrating how moving an application to the Lambda organization engineering decreased facilitating costs – by somewhere in the range of 66\% and 95\% and examines how further appropriation of this pattern may impact basic programming design configuration rehearses. Their first perception was on the effect of receiving a serverless design came amid the advancement of MindMup, a business online personality mapping application, which one of the creators is associated with creating and working. To research whether the impacts saw in the MindMup contextual analysis exchanged to different applications, they completed a contextual analysis of Yubl which is a London-based long-range interpersonal communication organization which gives Node.js back-end administrations to cell phone applications.

\section*{GPU enabled serverless computing frameworks}

With an increasing computing requirement for GPUs (Graphic Processing Units), more HPC (High Performance Computing) clusters are introducing high-end GPU-based accelerator. However, this poses a severe cost constraint instigating due an increase in the overall power consumption of each node. Several works have been conducted with an effort to reduce energy consumption in data centers \cite{b45}\cite{b46}\cite{b47}\cite{b48}\cite{b49}\cite{b50}. Though advancements have been made in GPU-based servers to make them power efficient than CPU-based servers, they tend to more expensive. Nvidia claims that their new Turing microarchitecture 25 times higher energy efficient and delivers 10 times higher performance than CPU-based servers \cite{b51}. 

However, costs are significant and GPUs supporting the new Turing architecture costs between US\$ 599 to US\$ 10,000 \cite{b52}. Though costs may reduce with economies of scale and due to increase in the demand and availability of GPUs. This is primarily due to the `gamer-driven' business volume as GPGPUs (General Purpose Graphical Processing Units) as they perform floating point operations, oversampling and interpolation quickly and efficiently \cite{b53}. Such calculations allow GPUs to render polygon shapes and texture mapping better than CPUs \cite{b54}. Furthermore, Turing-based GPU accelerators also provide support for deep learning via frameworks such as PLASTER \cite{b55}. However, in data preprocessing centers and HPC clusters, a single node with a high-end GPU may consume between 30~50\% more energy \cite{b44}. 

Virtualizing the GPU could contribute to better cost and power savings over a prolonged period. Though type-1 hypervisors or bare metal hypervisors such as Xen \cite{b56}, which is open sourced, can be used. They virtualize the entire system and may generate unnecessary overhead that is unacceptable for GPU driven applications. Other graphics APIs such as OpenGL \cite{b57} and Microsoft’s Direct 3D \cite{b58} facilitate the rendering of 2D and 3D vector graphics. Furthermore, they also achieve hardware-accelerated rendering \cite{b59} when interacting with the GPU. Another advantage is OpenGL’s and Direct 3D’s open, cross-language and cross-platform specification.

Chromium \cite{b60}, a flexible stream processing framework proposed by Humphreys et al, is one of many GPU virtualization applications/frameworks for graphics acceleration. Chromium manipulates streams of graphics API commands and enables OpenGL parallel rendering on clusters of workstations. This is done using wrapper libraries to make use of API interception and hence abstracting away underlying graphics architecture with a parallel interface. Thus, allowing it to also be a cross-application framework which can run on a range of different environments. 

Other OpenGL based rendering hardware approaches include VMGL \cite{b61}, which follows a similar approach to that of Chromium and provisions a VMM-Independent Graphics Acceleration. Furthermore, VMGL also supports the added capability of suspending and resuming across GPUs from different vendors. In addition to this, rendering performance in near native or within a 14\% window. Blink \cite{b62} is another similar approach that “safely multiplexes complex graphical content from a variety of VMs” while also supporting the emulation of legacy OpenGL programs.

rCUDA \cite{b44}, proposed by Duato et al, is a remote GPU acceleration framework for HPC clusters. Hence allowing multiple clients to remotely share GPUs. Furthermore, enabling GPU-based code acceleration remotely consequently reduces global power consumption for cluster nodes. This is a result of reducing the number of accelerators in each cluster node. Furthermore, drawbacks caused by the restrictive nature of Nvidia CUDA runtime API were addressed using a trivial solution, i.e., avoiding the use CUDE C extensions in the framework. 

Shi et al present a similar solution for GPU acceleration, termed vCUDA \cite{b63}. vCUDA encapsulates runtime APIs into RPC calls to achieve API interception and redirection. Thus, applications that utilize CUDA and execute within a VM can transparently access GPU hardware remotely. Hence allowing them to leverage hardware acceleration to significantly improve performances of HPC clusters, rather than directly incorporating a GPU in each independent cluster. To improve the efficiency of the RPC tool, frequency of procedure calls was reduced using shared memory mechanisms in VMM and lazy updates. An example of lazy remote procedure call and its implementation in C can be found in a paper by Feeley \cite{b64}. 

However, since vCUDA and rCUDA frameworks require to call and receive request from the remote CUDA APIs. In order to do this, users must install CUDA runtime wrapper library in their local environment and configure complex networks. Furthermore, any upgradation to a later version of CUDA would require users to reconfigure the entire CUDA runtime wrapper library for that particular version. Hence, such frameworks suffer from a complex build processes and several configuration processes and make them inconvenient to use. In addition to this, the added drawback of GPU communication overhead makes it unsatisfactory.

A novel GPU enabled serverless framework was proposed by Jun et al \cite{b65}. This solution allows developers to run deep learning programs remotely with minimal performance degradation and for clients who want to deploy micro-services requiring GPU support. This framework doesn’t require a CUDA or any GPU environment to be setup, eliminating the tedious configuration process. The architecture is based on iron.io’s IronFunctions \cite{b66} with some necessary enhancements. Hence, allowing the framework to incoporate Nvidia-docker via an API and IronFunctions. Thus, enabling users to utilize commands for GPU programming and also deploy Python-based high-performance services with the help of a python wrapper, PyCUDA \cite{b67}.

\section*{Serverless Computing: Economic and Architectural Impact}

Gojko Adzic and Robert Chatley \cite{b68} presents two case modern investigations of early adopters, demonstrating how moving an application to the Lambda organization engineering decreased facilitating costs – by somewhere in the range of 66\% and 95\% and examines how further appropriation of this pattern may impact basic programming design configuration rehearses. Their first perception was on the effect of receiving a serverless design came amid the advancement of MindMup, a business online personality mapping application, which one of the creators is associated with creating and working. To research whether the impacts saw in the MindMup contextual analysis exchanged to different applications, they completed a contextual analysis of Yubl which is a London-based long range interpersonal communication organization which gives Node.js back-end administrations to cell phone applications.

\section*{Serving deep learning models in a serverless platform}

Vatche Ishakian and Vinod Muthusamy \cite{b69} assess the appropriateness of a serverless figuring condition for the inferencing of vast neural system models. Their test assessments are executed on the AWS Lambda condition utilizing the MxNet profound learning system. Their test results demonstrate that while the inferencing idleness can be inside an adequate range, longer postponements because of virus begins can skew the idleness conveyance and thus chance abusing more stringent SLAs.One of the objectives of their assessment was to comprehend if a serverless figuring stage can be used for neural system inferencing. The trials utilized the AWS Lambda serverless stage, also, the Amazon MXNet profound learning system. In the future, they intend to stretch out their assessments to incorporate other serverless stages and profound learning systems, for example, Tensorflow among others.

\section*{Pay-Per-Request Deployment of Neural Network Models Using Serverless Architectures}

Zhucheng Tu, Mengping Li, and Jimmy Lin \cite{b70} shows the serverless arrangement of neural systems for model inferencing in NLP applications utilizing Amazon's Lambda administration for feedforward assessment and DynamoDB for putting away word embeddings. Their design understands a compensation for every solicitation estimating model, requiring zero continuous expenses for keeping up server occasions. All virtual machine the board is taken care of in the background by the cloud supplier with no immediate designer intercession. They portray various strategies that permit proficient utilization of serverless assets, what's more, assessments affirm that their plan is both versatile and cheap.

In this work, they chose Amazon Web Services (AWS) as their arrangement stage because of its showcase overwhelming position, albeit other cloud suppliers have comparable contributions. To empower serverless neural system induction for NLP, the prepared models are bundled together with the capacity to be conjured and subordinate programming libraries. The cloud supplier is in charge of making the earth for execution. Amid induction, the Lambda work takes input content, which is provided remotely by means of an API portal. Sentences should be first changed into an implanting lattice built utilizing word vectors. These are gotten from DynamoDB. At long last, the Lambda work applies feedforward assessment on the installing framework as indicated by the provided display, yielding a last forecast. 

\section*{MXNet: A Flexible and Efficient Machine Learning Library for Heterogeneous Distributed Systems}

Tianqi Chen, Mu Li and Yutian Li \cite{b71} talks about which is a MXNet is a multi- language AI (ML) library to facilitate the improvement of ML calculations, particularly for profound neural systems. Inserted in the host language, it mixes definitive emblematic articulation with basic tensor calculation. It offers auto separation to infer slopes. MXNet is calculation and memory productive and keeps running on different heterogeneous frameworks, going from cell phones to disseminated GPU groups. This paper depicts both the API structure and the framework execution of MXNet, and clarifies how installing of both emblematic articulation and tensor task is dealt with in a brought together manner. Our fundamental tests uncover promising outcomes on expansive scale profound neural system applications utilizing numerous GPU machines.

\section*{Acknowledgment}

The authors would like to thank Dr. Jothi KR for his continuous support throughout this paper. I would also like to thank Vellore Institute of Technology for their aid without which this paper wouldn't have been completed.

\end{document}